\documentstyle[11pt,newpasp,twoside,epsf,graphicx]{article}
\def\edcomment#1{\iffalse\marginpar{\raggedright\sl#1\/}\else\relax\fi}
\reversemarginpar

\begin{document}
\title{A High Angular Resolution View of Hot Gas in Clusters, Groups,
and Galaxies -- Mergers, Mixing, and
Bubbling}
\author{W. Forman$^1$, E. Churazov$^{2,3}$, L. David$^1$,
F. Durret$^4$, C. Jones$^1$, M. Markevitch$^1$, S. Murray$^1$,
M. Sun$^1$, A. Vikhlinin$^{1,3}$}
\affil{$^1$Smithsonian Astrophysical Observatory, Cambridge,
MA, USA\newline
$^2$MPI f\"{u}r Astrophysik, 85740 Garching, Germany\newline
$^3$Space Research Institute, 117997 Moscow, Russia\newline
$^4$Institut d'Astrophysique, 98bis Boulevard Arago, Paris, France}

\begin{abstract}

We discuss two themes from Chandra and XMM-Newton observations of
galaxies, groups, and clusters. First, we review observational aspects
of cluster formation and evolution as matter accretes along filaments
in A85 and A1367.  We describe Chandra observations that probe the later
evolutionary phases where the effects of mergers -- both subsonic and
supersonic -- are observed in cluster cores as ``cold fronts'' and
shocks. Second, we review the interactions between the hot,
intracluster gas with relativistic plasma originating in active nuclei
within the dominant galaxy at the cluster center. As examples of this
interaction, we describe the radio and X-ray observations of M87 where
buoyantly rising bubbles transfer energy and matter within the cluster
core. We describe the Chandra observations of ZW3146 which exhibits
both multiple cold fronts and relativistic plasma interactions. Finally, we
describe the X-ray observations of NGC4636 where energy produced
by the central AGN imprints a unique signature on the surrounding hot
corona of the galaxy.
\end{abstract}

\section{Cluster Evolution and Mergers}

The most popular view of the formation of clusters, the largest
gravitationally collapsed systems in the Universe, is that they form
hierarchically from smaller mass systems at the nodes of large scale
filaments.  At the present epoch, matter continues to accrete onto
clusters preferentially along the large scale filaments.  X-ray
observations of present epoch clusters, combined with optical studies,
show both the large scale merging process as well as provide new
insights into the detailed physics as merger remnants traverse cluster
cores.

\subsection{Filamentary Structures and Cluster Formation}

A85 shows the effects of filamentary structure on the appearance of
the cluster. In a remarkable ROSAT observation (Durret et al. 1998), a
large scale X-ray filament was detected extending from a merging
subcluster (South Blob) towards the southeast (see Fig.~1a) at a
position angle of $\sim160^{\circ}$ (measured from north,
counterclockwise). The major axis of the central cD galaxy, the
distribution of bright cluster galaxies, the X-ray filament and nearby
groups and clusters all show a common orientation at this same
$\sim160^{\circ}$ angle over linear scales extending from 100 kpc (the
outer isophotes of the central cD galaxy) to 25 Mpc, the alignment of
nearby clusters (Durret et al. 1998). Such common alignments over a
wide range of scales are expected if clusters form through accretion
of matter from filaments and the accretion direction remains fixed
over significantly long fractions of the age of the Universe (e.g.,
Van Haarlem \& Van de Weygaert 1993).  Recent XMM-Newton observations
of A85 detect X-ray emission from the inner region of the filament
extending from the southern clump ($10'$ south of the X-ray peak; see
Fig.~1b) for about $15'$ to the southeast (Durret et al. 2003; see
Kempner et al. 2002 for a discussion of the Chandra observation of the
South Blob).

\begin{figure}
\centerline{\includegraphics[width=0.5\linewidth]{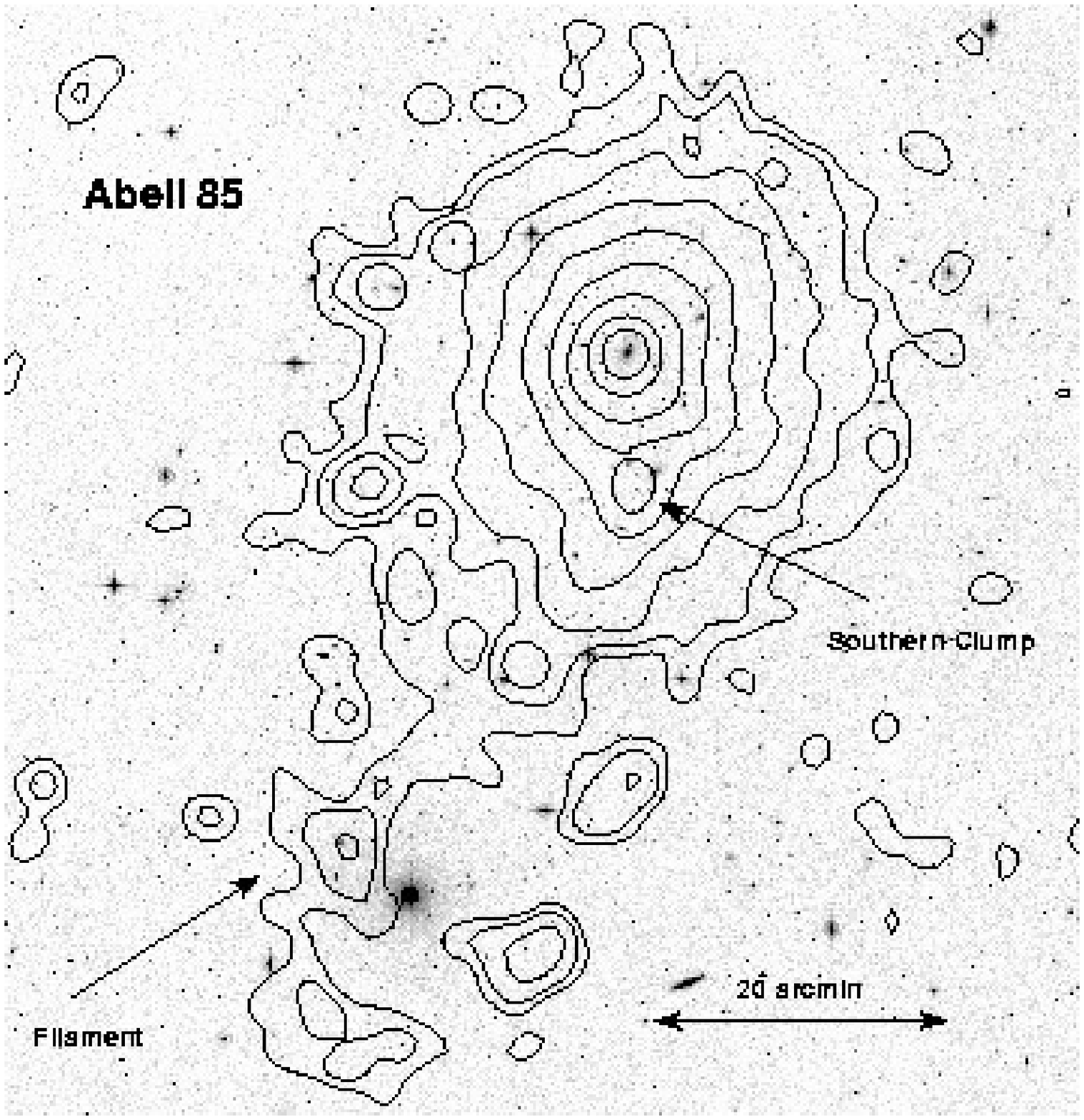}
\includegraphics[width=0.5\linewidth]{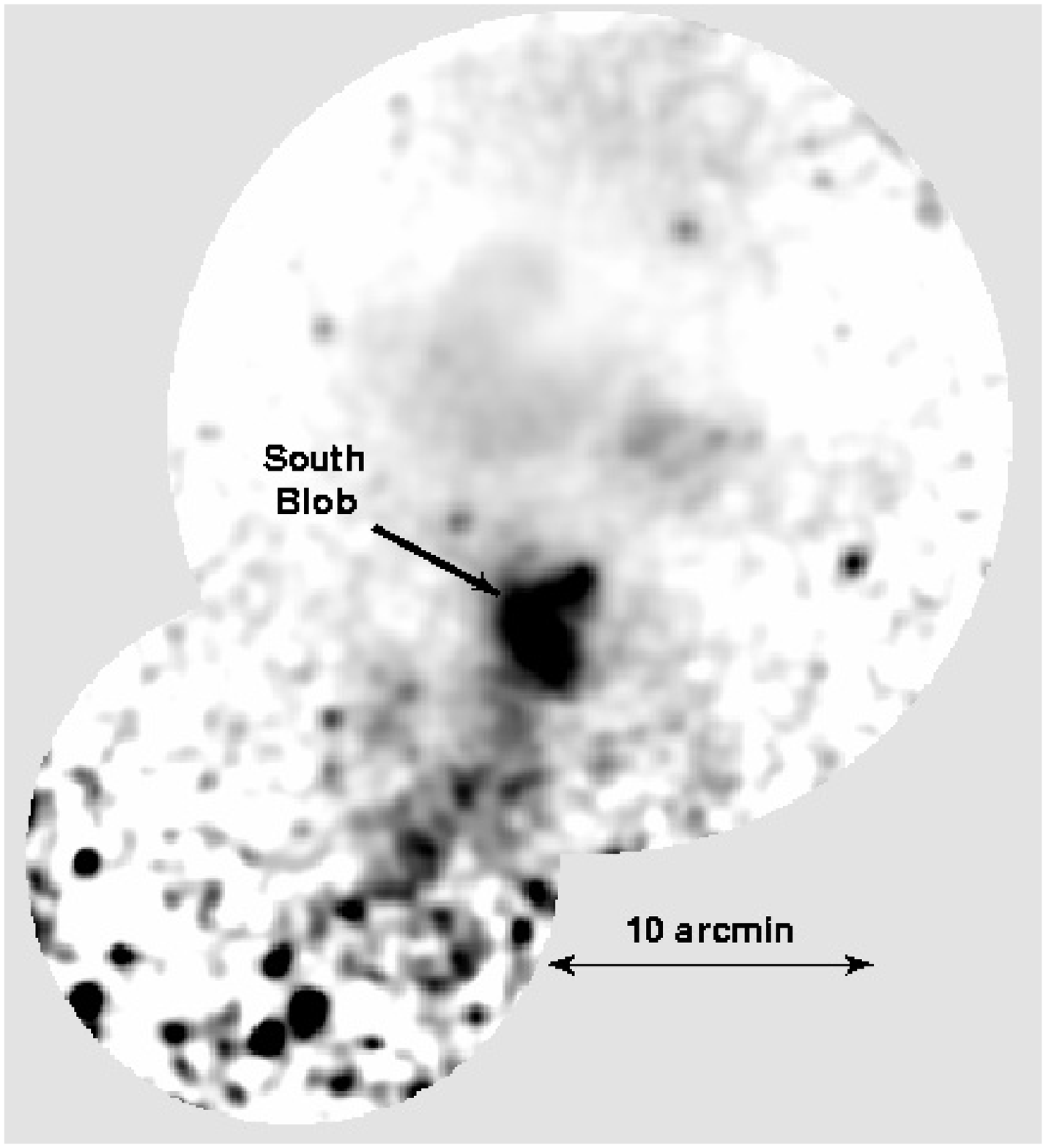}}
\caption{ROSAT and XMM-Newton observations of A85 (\textbf{a}) 
ROSAT PSPC iso-intensity contours (0.4-2.0 keV) are shown superposed on
an optical image (adapted from Durret et al. 1998). A filamentary
structure extends to the southeast.
(\textbf{b}) The XMM-Newton MOS image shows the inner portion of the
filament extending southeast from the South Blob (Southern Clump) (Durret et al. 2003).
}
\end{figure}

A second example of a cluster that clearly shows the importance of the
surrounding filamentary structure is A1367. As shown by West \&
Blakeslee (2000), A1367 lies at the intersection of two filaments --
the first extending roughly 100~Mpc from A1367 towards Virgo and the second extending
between A1367 and Coma.  The X-ray structure of A1367 as seen in the
XMM-Newton observation in Fig.~2a and 2b shows effects from both these
filaments with cool gas streaming into the cluster core from the
direction of Coma to the northeast and a merger underway along the
axis towards Virgo along the northwest--southeast axis (Forman et
al. 2003a; see additional details on A1367 in the contribution by
M. Sun in these proceedings).

A85 and A1367 contrast the effects of accretion. For A85, matter is
accreting from a filament into a well-developed ``cooling flow''
cluster with a low entropy core centered on a cD galaxy. For A1367, we
see accreting gas penetrating directly into the core of a less relaxed
cluster.

\begin{figure}
\centerline{
\includegraphics[width=0.3\linewidth,bb=190 260 430 490,clip]{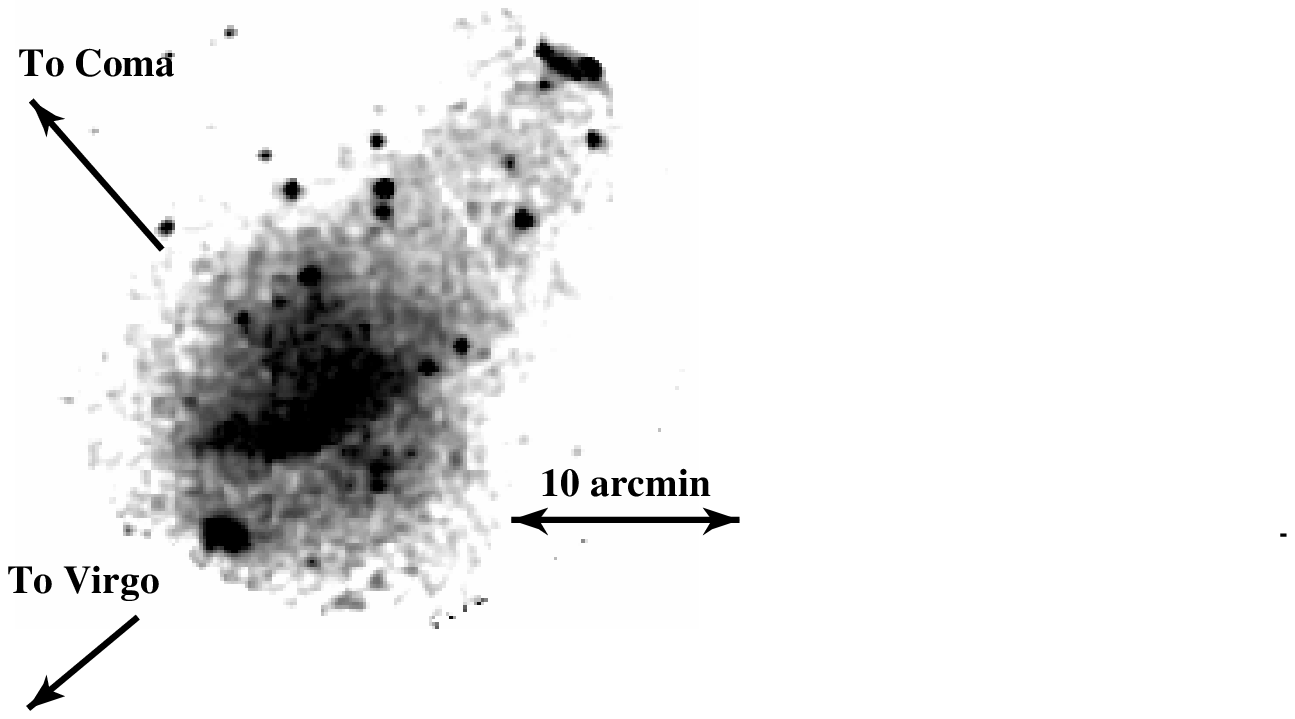}
\includegraphics[width=0.3\linewidth,bb=140 240 455 525,clip]{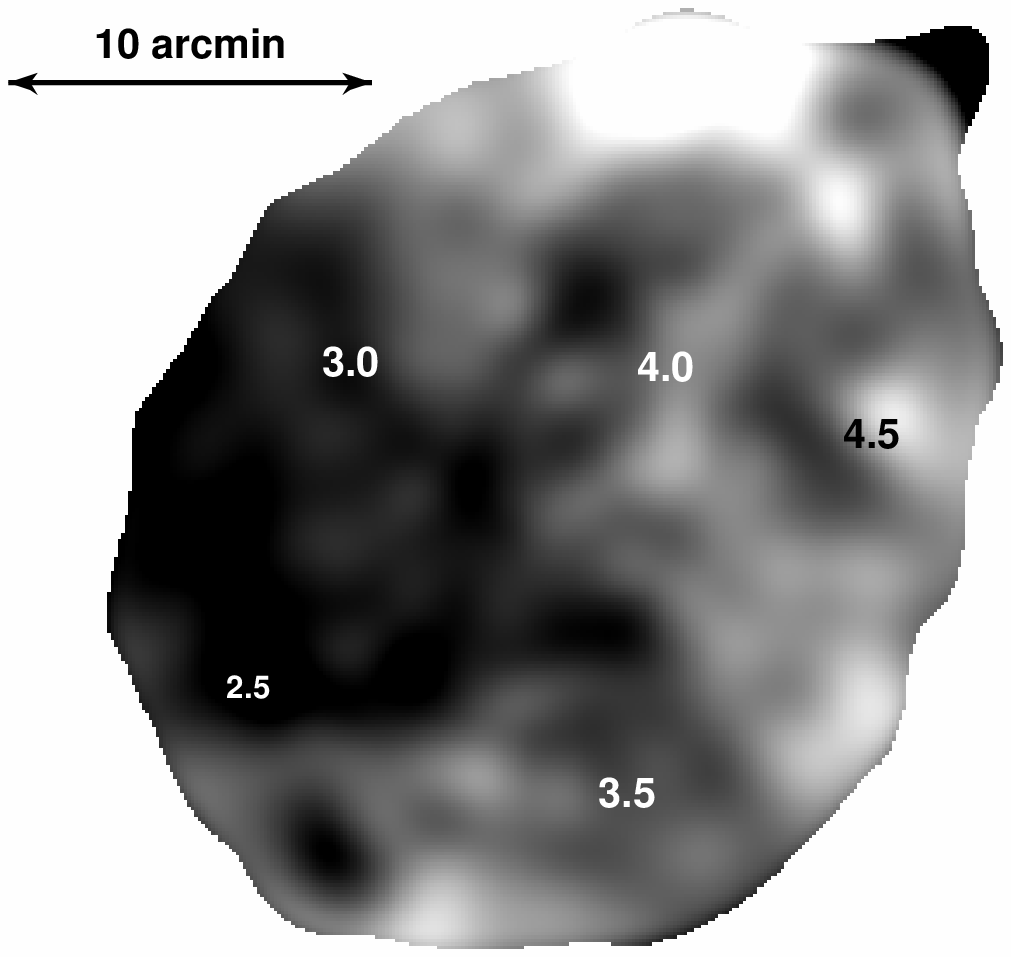}
\includegraphics[width=0.3\linewidth,bb=20 225 560 690,clip]{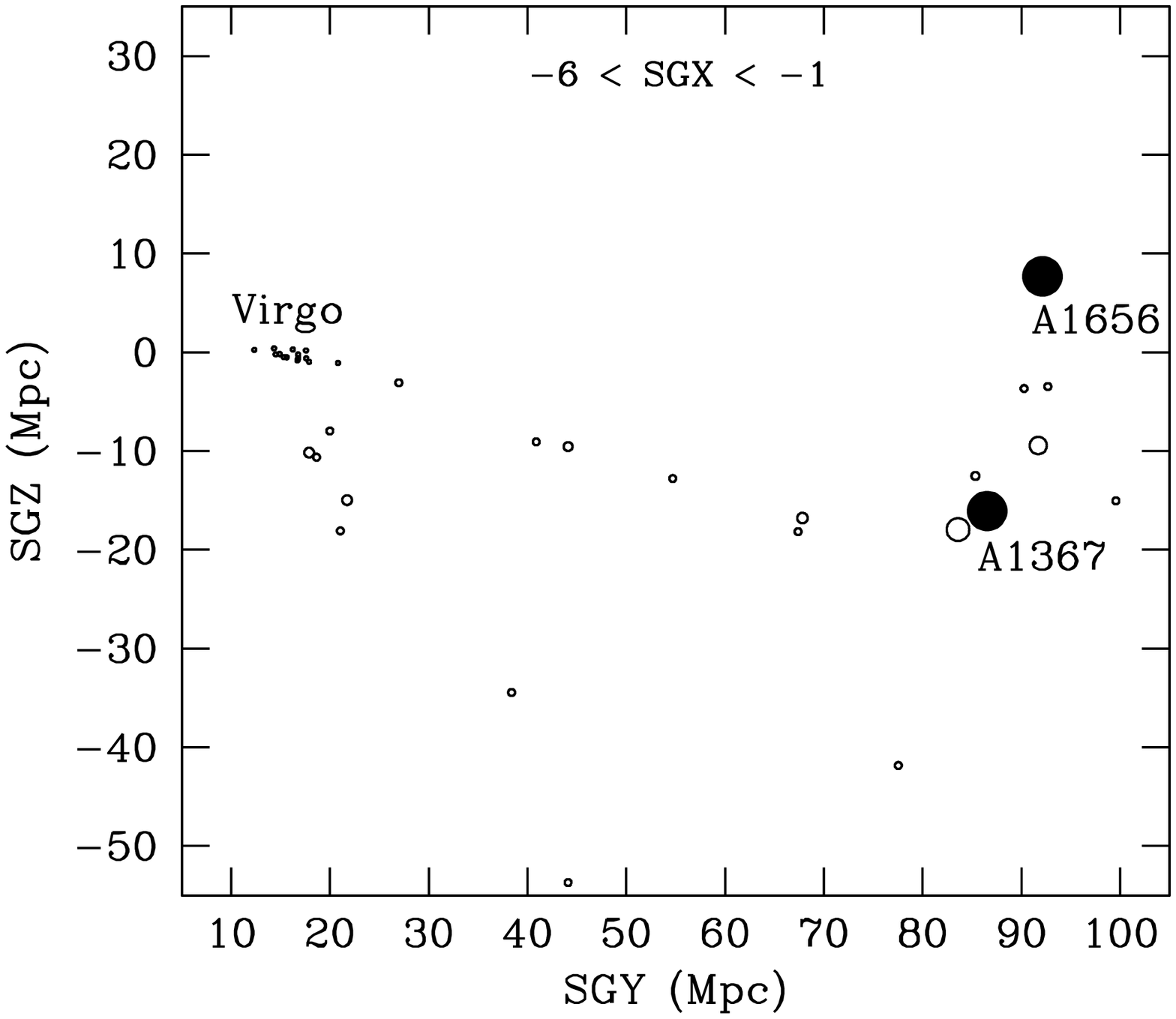}
}
\caption{XMM-Newton observations of A1367 (Forman et al. 2003a) 
(\textbf{a}) Smoothed, flat-fielded, background-subtracted XMM-Newton (MOS)
image shows the remarkable ellipticity of A1367 with the extension to
the northwest. Also seen is the bright, highly structured
core. (\textbf{b})  The temperature map (XMM-Newton MOS data) shows the
extensive cool gas (darker shading is cooler gas; numbers indicate the
gas temperature for the corresponding grey scale) entering the cluster
from the east. (\textbf{c}) The large scale structure surrounding A1367
(adapted from West \& Blakeslee 2000) shows filaments extending from A1367
towards the Virgo and Coma clusters.}
\end{figure}

\subsection{Cold Fronts}

The study of ``cold fronts'', contact discontinuities between cooler
and hotter gas, began with the launch of Chandra (Markevitch et
al. 2000; Vikhlinin et al. 2001a, b).
A particularly revealing example of a cold front is seen in the
Chandra observation of the Fornax cluster (Dosaj et al. 2002).  In
Fig.~3a, we see gas bound to the infalling bright
elliptical galaxy NGC1404 as it approaches the cluster center (to the
northwest). The image clearly shows the sharp edge of the surface
brightness discontinuity, shaped by the ram pressure of the Fornax cluster
gas.  The temperature map (Fig.~3b) confirms that the
infalling cloud is cold compared to the surrounding Fornax ICM.

\begin{figure}
\centerline{
\includegraphics[width=0.4\linewidth,bb=100 180 510 610,clip]{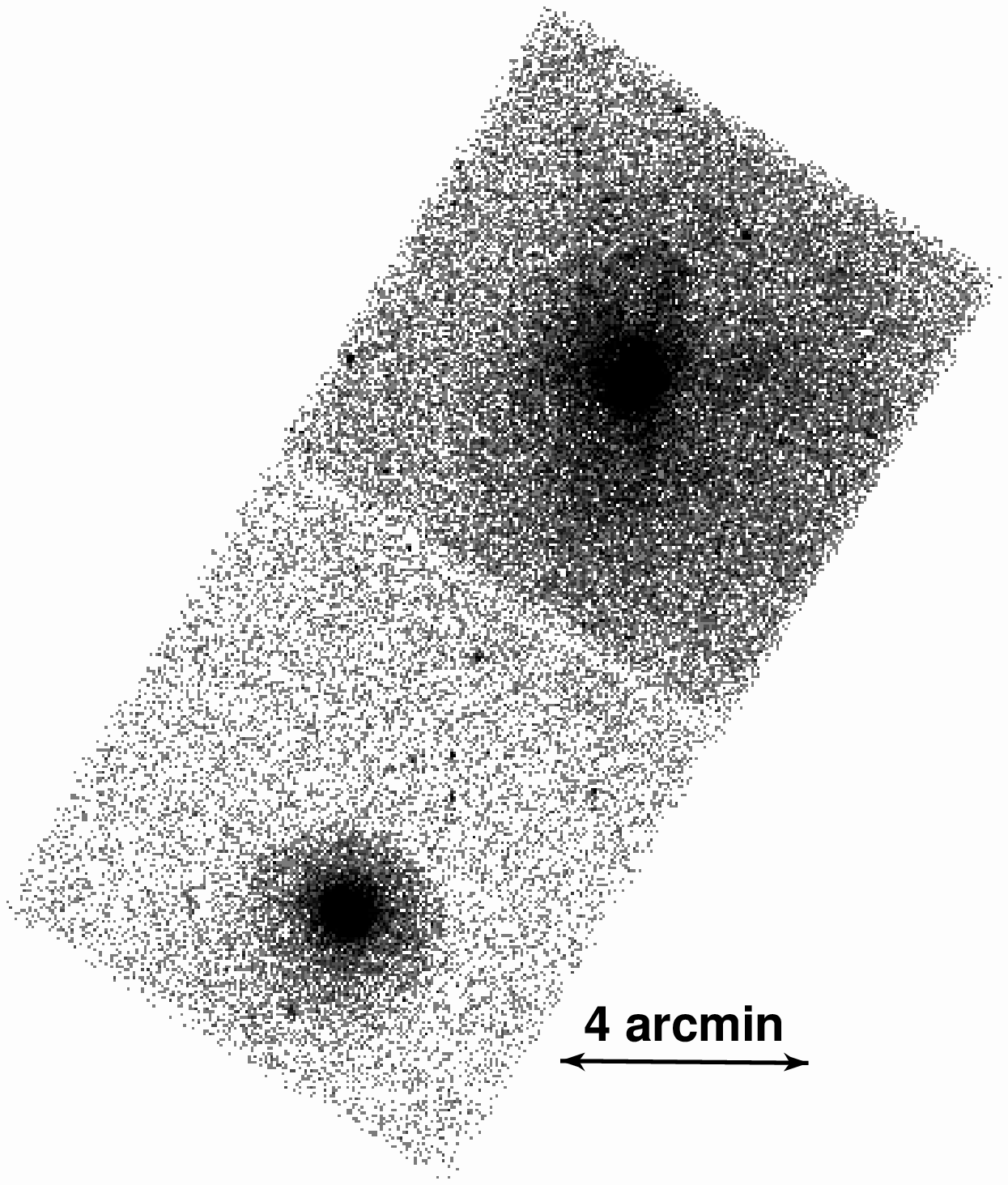}
\includegraphics[width=0.4\linewidth,bb=125 200 450 550,clip]{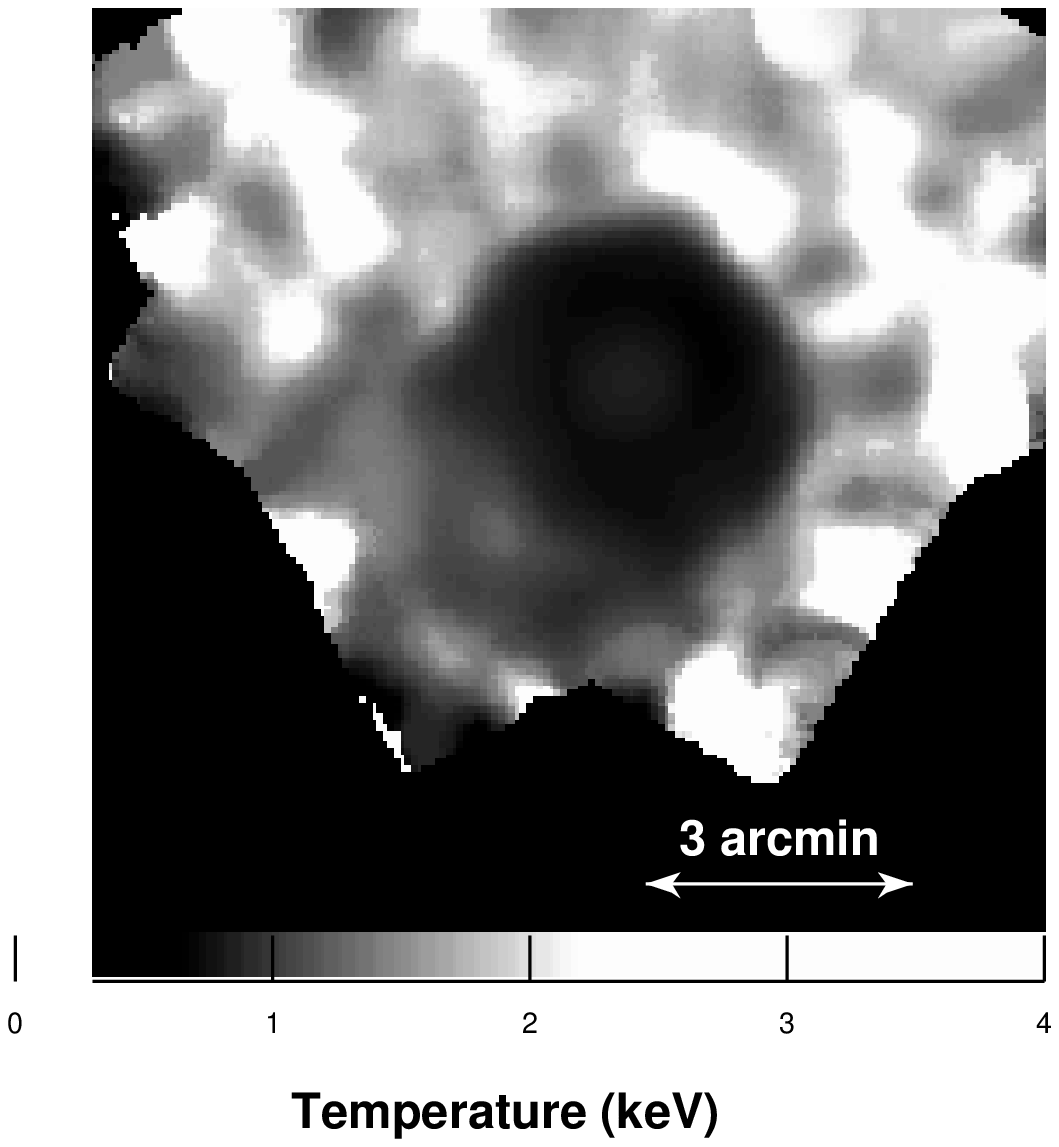}
}
\caption{The ACIS observation of NGC1404 and NGC1399. (\textbf{a}) The
0.5--2.0 keV band image of the Fornax cluster.  The gas filled dark
halo surrounding NGC1404 is at the lower left (southeast) while the
cluster core, dominated by the halo surrounding NGC1399 lies at the
upper right (northwest). (\textbf{b}) The temperature map of the
Fornax region. The cold core surrounding NGC1404 has a temperature of
less $\sim1$ keV while the surrounding gas has a temperature of
$_>\atop^{\sim}$1.5 keV.  }
\end{figure}

Cold fronts provide a unique opportunity to explore cluster physics.
In a study of A3667, Vikhlinin et al. (2001a, b) derived the ram
pressure of the ICM on the moving cold front from the gas density and
gas temperature. In turn, the ram pressure yielded a measurement of
the cold front velocity.  The factor of two difference in pressures
between the free streaming region and the region immediately inside
the cold front implied a cloud velocity of $1430\pm290$ km s$^{-1}$
(Mach $1\pm0.2$). In addition, Vikhlinin et al. (2001b) showed that
the ``edge'' of the cold front in A3667 is very sharp -- the width of
the front was less than $3.5''$ (5 kpc). This sharp edge requires that
transport processes across the edge be suppressed, presumably by
magnetic fields. Without such suppression, the density
discontinuity at the ``edge'' would be broader since the relevant
Coulomb mean free path for electrons is several times the width of the
cold front.  Furthermore, Vikhlinin et al. observed that the cold
front appears sharp only over a sector of about $\pm30^{\circ}$
centered on the direction of motion, while at larger angles, the sharp
boundary disappears. The disappearance can be explained by the onset
of Kelvin-Helmholtz instabilities, as the ambient ICM gas flows past
the moving cold front. To suppress the instability over the inner
$\pm30^{\circ}$ requires a magnetic field parallel to the boundary
with a strength of $7-16\mu$G (Vikhlinin et al. 2001b).

\subsection{A Classic Supersonic Merger - 1E0657}

1E0657 ($z=0.296$) was discovered by Tucker et al. (1995) as part of a
search for ``failed'' clusters, clusters that were X-ray bright but
had few, if any, optical galaxies. The cluster temperature, as
measured from ASCA observations, was found to be remarkably high,
about 17 keV, making it the hottest cluster known
(Tucker et al. 1998). 

The Chandra image of 1E0657 shows the classic
properties of a supersonic merger (see Markevitch et al. 2002a for a
detailed discussion).  The image shows a dense (cold) core moving to the west
after having traversed, and disrupted, the core of the main
cluster. Leading the cold, dense core is a density discontinuity that
appears as a shock front (Mach cone). The shock is confirmed by the spectral
data since the gas to the east (trailing the shock) is hotter than
that in front of the discontinuity, unlike the cold
fronts discussed above (or the western boundary of the bullet which
also is a cold front). The detailed gas density parameters confirm
that the ``bullet'' is moving to the west with a velocity of
3000--4000~km~sec$^{-1}$, approximately 2-3 times the sound speed of
the ambient gas. 1E0657 is the first, and so far the only,  clear example of a relatively
strong shock arising from cluster mergers.

\section{Interaction between Hot Gas in Galaxies, Groups, and Clusters with
Energy Produced by AGN Accretion}

With the greatly increased observing capabilities provided by Chandra
and XMM-Newton, the complexity of the phenomena exhibited by hot gas
in early type galaxies, groups and clusters seen in X-rays has grown.
Not only do we see the effects of merging, but also we see the detailed effects
on the hot gaseous atmospheres of the liberation of accretion energy
as relativistic plasmas are expelled from AGN.

One of the first, and clearest, examples of the effect of plasma
bubbles on the hot intracluster medium was found in the Perseus
cluster around the bright active, central galaxy NGC1275 (3C84). First
studied in ROSAT images (Bohringer et al. 1993), the radio emitting cavities to
the north and south of NGC1275 are clearly seen in the Chandra images
with bright X-ray emitting rims surrounding the cavities that coincide
with the inner radio lobes (Fabian et al. 2000). For NGC1275/Perseus, the
radio lobes are in approximate pressure equilibrium with the ambient,
denser and cooler gas and the bright X-ray rims surrounding the
cavities are cooler than the ambient gas.  The central galaxy in the
Hydra A cluster also harbors X-ray cavities associated with radio
lobes that also show no evidence for shock
heating (McNamara et al. 2000).
Both sets of radio bubbles, being of lower density than the ambient
gas, must be buoyant.

The Chandra images of Perseus/NGC1275 also suggest the presence of
older bubbles produced by earlier outbursts (Fabian et al. 2000). These
older bubbles appear as X-ray surface brightness ``holes'', but unlike
the inner bubbles, these outer holes show no detectable radio
emission, suggesting that the synchrotron emitting electrons may have
decayed away leaving a heated, plasma bubble (see Fabian et al. 2002a
who reported low frequency radio spurs extending towards the
outer bubbles in NGC1275, consistent with this
scenario).  Such bubbles, with no attendant radio
emission, are seen by Chandra in the galaxy groups HCG62 and
MKW3s (Vrtilek et al. 2001; Mazzotta et al. 2002).

In addition to bubbles associated with
central dominant cluster galaxies, bubbles and their effects
also are seen in more typical early type galaxies.  For example, in the E1
galaxy M84 (NGC4374), Chandra observed an unusual X-ray morphology
which is explained by the effect of the radio lobes on the hot
gas (Finoguenov \& Jones 2001). The X-ray emission appears ${\cal
  H}$-shaped, with a bar extending east-west with two nearly parallel
filaments perpendicular to this bar.  The complex X-ray surface
brightness distribution arises from the presence of two radio lobes
(approximately north and south of the galaxy) that produce two low
density regions surrounded by higher density X-ray filaments. As with
Perseus/NGC1275 and Hydra A, the filaments, defining the ${\cal
  H}$-shaped emission, have gas temperatures comparable to the gas in
the central and outer regions of the galaxy and hence argue against
any strong shock heating of the galaxy atmosphere by the radio plasma.

Although the bubbles are not driving shocks into the surrounding gas,
they still can provide significant energy input. One particularly
well-studied system combining complex X-ray and radio emission is M87,
at the center of the Virgo cluster.  The 327 MHz high resolution, high
dynamic range radio map of M87 (see Fig.~4a) shows a well-defined
torus-like eastern bubble and a less well-defined western bubble, both
of which are connected to the central emission by a column, and two
very faint almost circular emission regions northeast and southwest of
the center (Owen et al. (2000). The correlation between X--ray and
radio emitting features has been remarked by Feigelson et al. (1987),
Bohringer et al. (1995), and Harris et al. (1999).

\begin{figure}
\plottwo{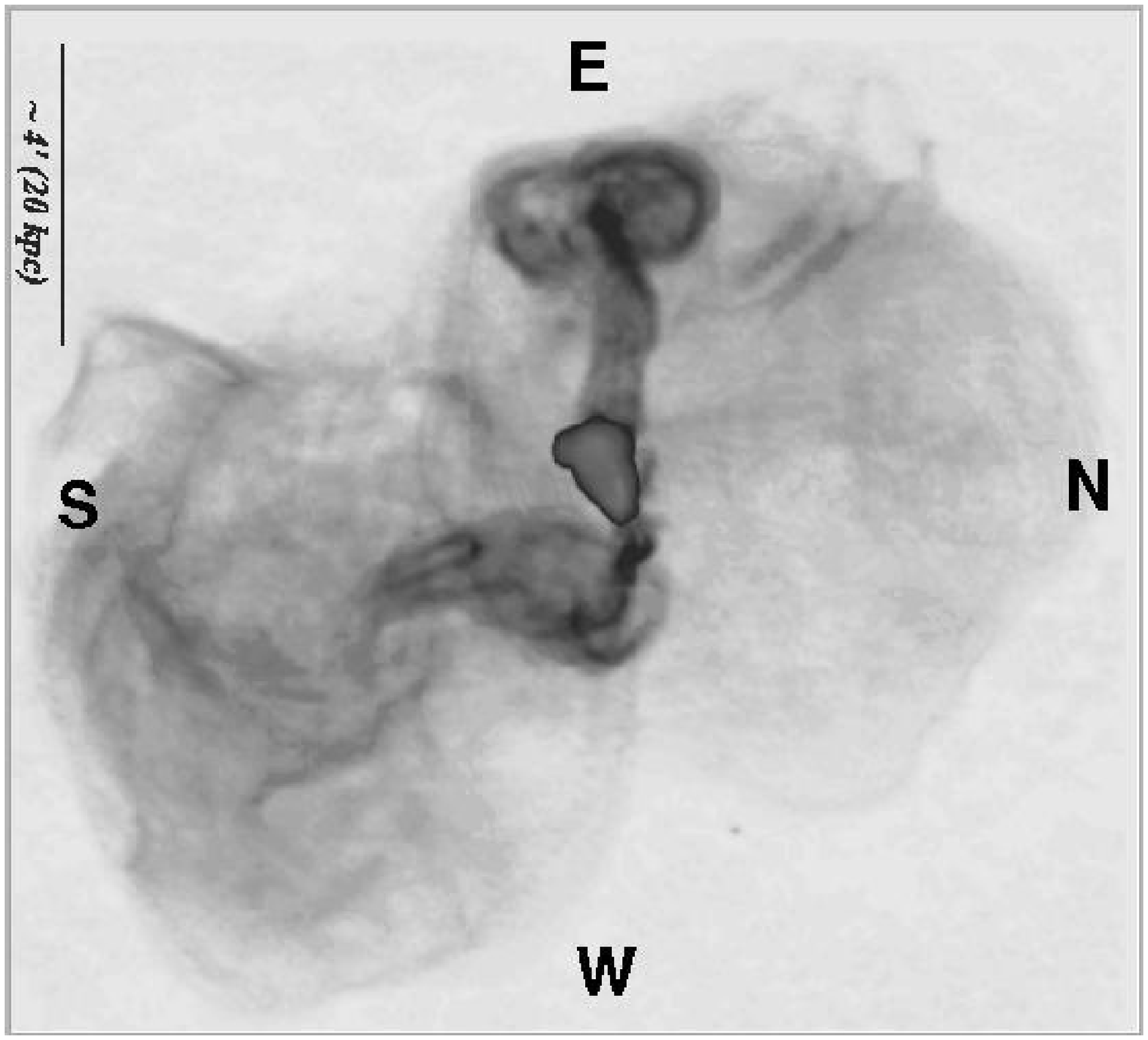}{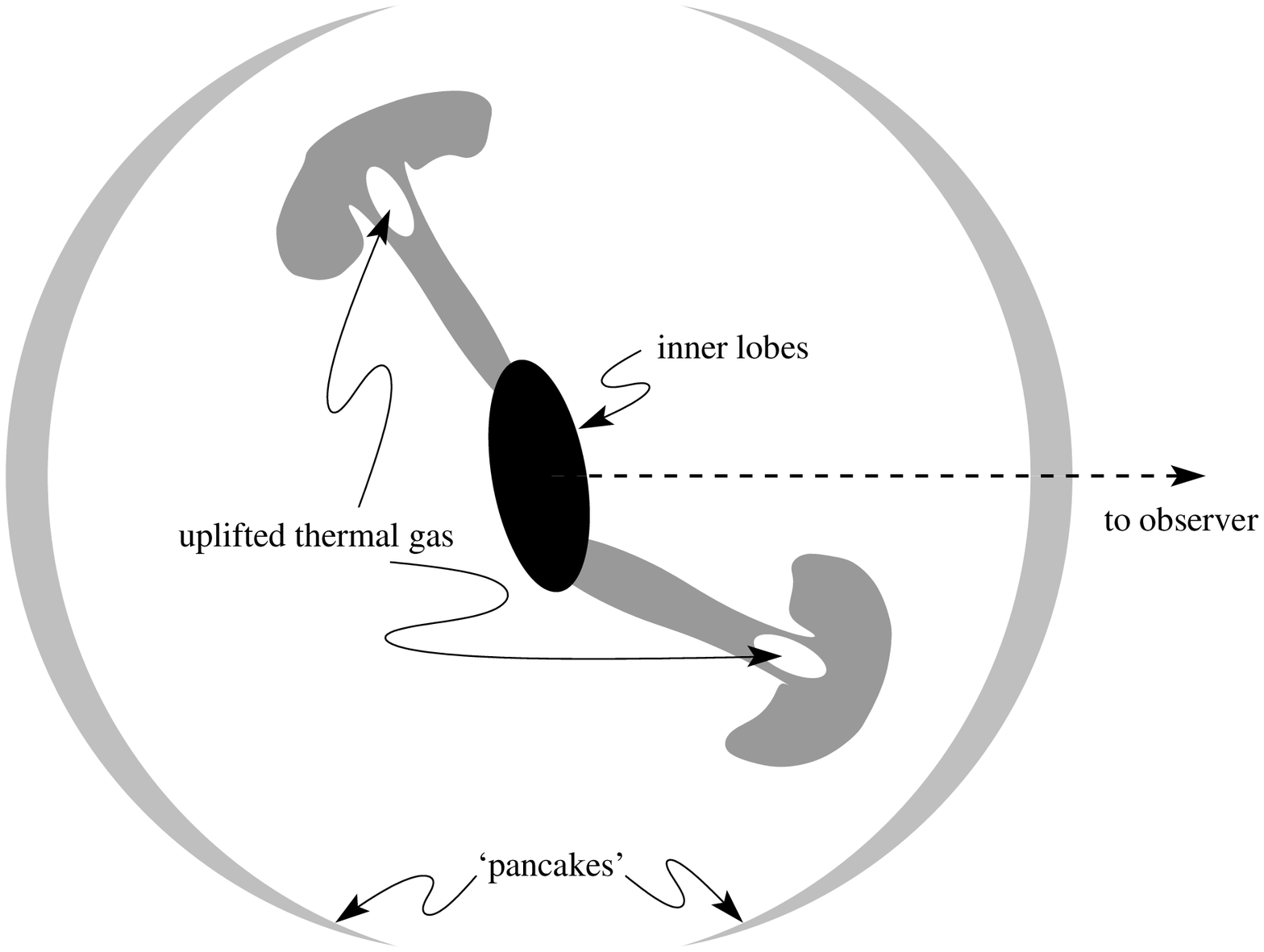}
\caption{ (\textbf{a})
$14'.6 \times 16'.0$ radio map of M87 (North to
  the right, East is up) (from Owen et al. 2000).
(\textbf{b}) Source
  geometry.  The central black region represents the inner radio lobes,
  the gray ``mushrooms'' correspond to buoyant bubbles,
  transformed into tori, and the gray lens-shaped structures are
  ``pancakes'' (seen edge-on) formed by older bubbles. }
\end{figure}

Motivated by the similarity in appearance between M87 and hot bubbles
rising in a gaseous atmosphere, Churazov et al. (2000) developed a
simple model of the M87 bubbles which is generally applicable to the
many bubble-like systems seen in the Chandra observations.  An initial
buoyant, spherical bubble transforms into a torus as it rises through
the galaxy or cluster atmosphere.  By entraining cool gas as it rises,
it exhibits a characteristic ``mushroom'' appearance, similar to an
atmospheric nuclear explosion.  This may qualitatively explain the
correlation of the radio and X--ray emitting plasmas and naturally
accounts for the thermal nature of the X-ray emission associated with
the rising torus (Bohringer et al. 1995).  Ambient gas is uplifted in
the cluster atmosphere producing the ``stem'' of the mushroom that is
brighter than the surrounding regions (Churazov et al. 2000).
Finally, in the last evolutionary phase, the bubble reaches a height
at which the ambient gas density equals that of the bubble.  The
bubble then expands to form a thin layer (a ``pancake'').  The large
low surface brightness features in the M87 radio map could be just
such pancakes (see Fig.~4b for a schematic of the radio emitting
components of M87).

The XMM-Newton observation (see Belsole et al. 2001) supports the
buoyancy scenario with spectra showing that emission from the X-ray
columns is thermal in nature with a gas temperature that is lower (1.5
keV) than the surrounding gas (2.3 keV).

\subsection{ZW3146 - Multiple Edges}

\begin{figure} [t]
\centerline{
\includegraphics[width=0.45\linewidth,bb=160 250 520 520,clip]{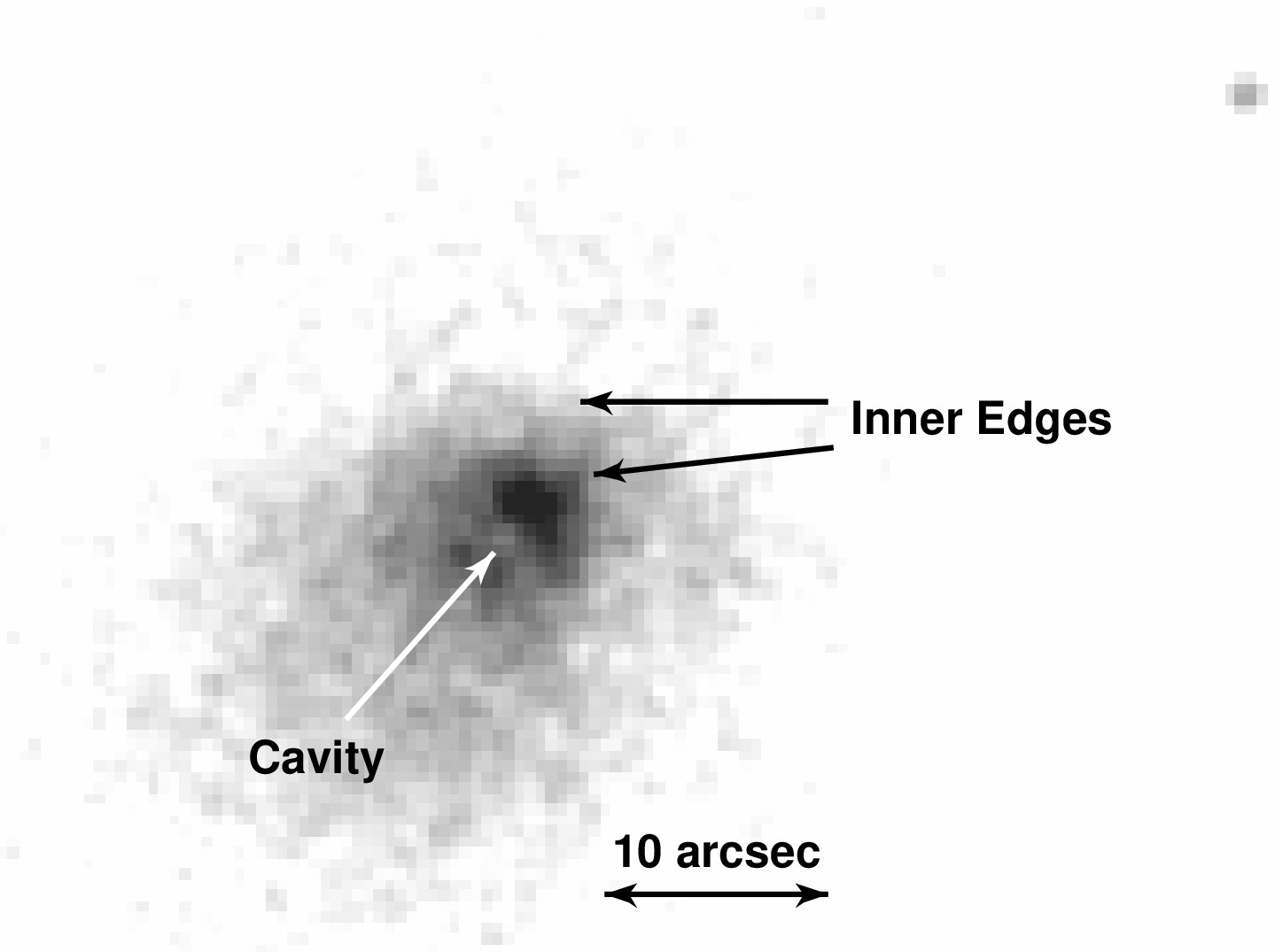}
\includegraphics[width=0.45\linewidth,bb=50 250 500 575,clip]{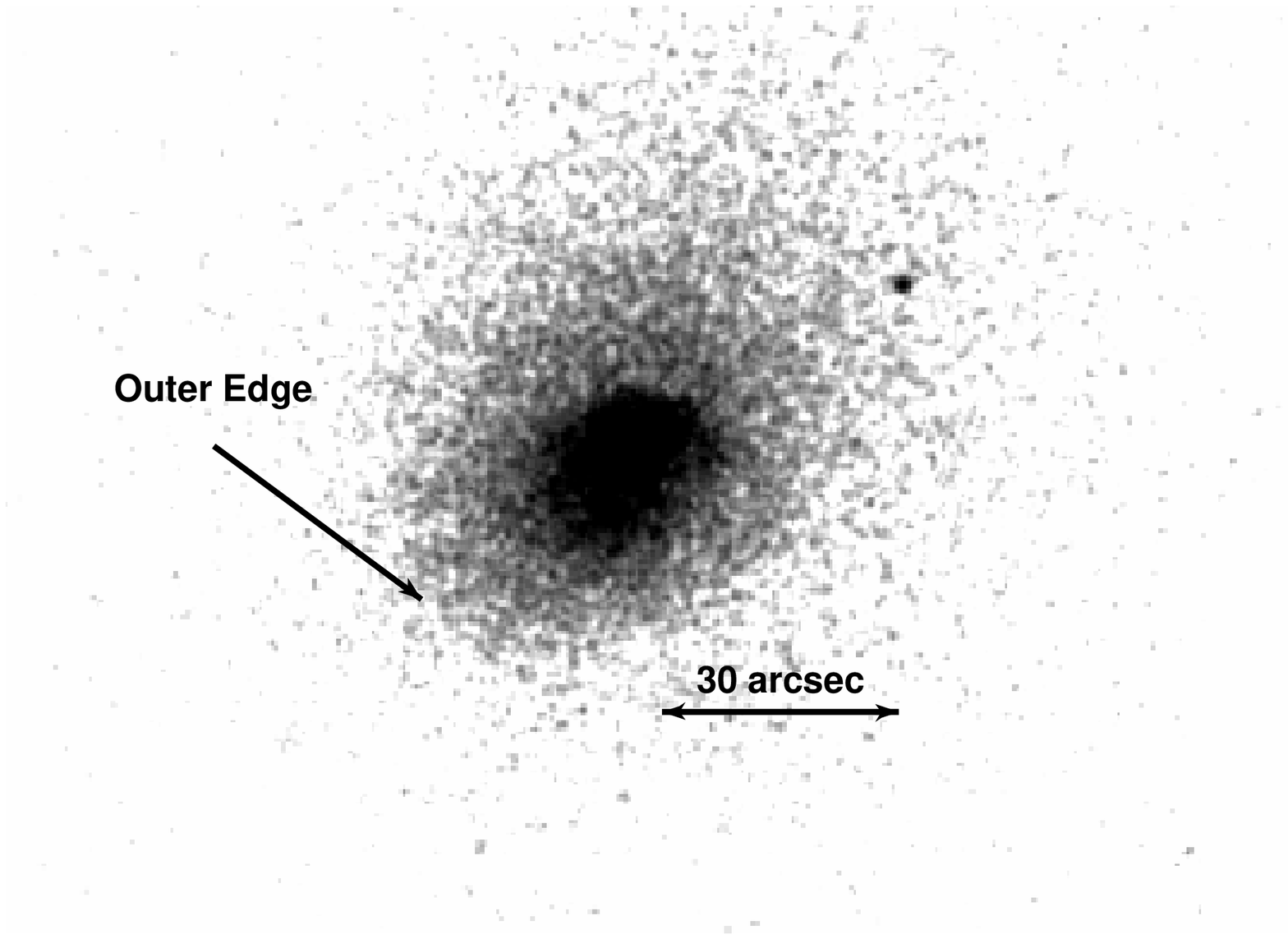}
}
\caption{Edges in ZW3146. (\textbf{a}) The 0.5-2.0 keV image of the central
  region of ZW3146 shows two ``edges'' as well as an inner cavity at
the position of a nuclear radio source which lies at the center of the
optical image of the cD galaxy. (\textbf{b}) A third ``edge'' appears
on a larger scale at $\sim35''$ from the cD galaxy center. }
\end{figure}

ZW3146 is a moderately distant ($z=0.29$; 5.74 kpc per arcsec)
cluster with a remarkably high mass deposition rate that has been
estimated to exceed 1000 M$_\odot$~yr$^{-1}$ (Edge et al. 1994). The
Chandra image further demonstrates the remarkable nature of this
cluster -- on scales from $3''$ to $30''$ ($\sim20$ kpc to 170 kpc),
three separate ``edges'' are detected (see Fig.~5 and Forman et
al. 2003b). At the smallest radii, two edges are seen to the northwest
and north (see Fig.~5a).  The first, at a radius of $\sim3''$ (17 kpc)
has a surface brightness drop of almost a factor of 2. This innermost
``edge'' defines a cavity, surrounded by a partial shell, centered on
the bright, central cD galaxy. The cavity probably arises as
relativistic plasma (seen as radio emission in the NVSS/FIRST survey)
produced in the central AGN expands outward and evacuates the X-ray
emitting plasma from the immediate environs of the the nucleus.  The
second edge, at a radius of $\sim8''$ (45 kpc) has a surface
brightness drop of almost a factor of 4.  The third ``edge'' (Fig.~5b)
lies to the southeast, about $35''$ (200 kpc) from the cluster center.
One possibility for the origin of these multiple edges is that the
$8''$ edge is produced by gas motions induced by the expanding plasma
bubble in the central core. Alternatively, ``sloshing'' motions can
arise either from mergers or bubbles (e.g., Markevitch et al. 2001;
Churazov et al. 2003).

Studies of clusters like ZW3146, for which the standard cooling flow
scenario predicts large mass deposition rates but which have complex
cores, may help to resolve the source of the energy needed to moderate
the effects of radiative cooling as required by recent XMM and Chandra
observations (e.g., Peterson et al. 2001 and Tamura et al. 2001).  A
variety of proposals have been made to explain the much smaller
amounts of cool gas than expected (e.g., Churazov et al. 2001, Bruggen
\& Kaiser 2001, Quilis et al. 2001, Bohringer et al. 2002, Nulsen et
al. 2002, Fabian et al. 2002b, 2002c, Ruszkowski \& Begelman 2002,
Zakamska \& Narayan 2002, Churazov et al. 2002).  Only more
observational studies can determine which processes are most
effective.

\subsection{Origin of Density Edges}

The variety of morphologies and scales exhibited by sharp edges or
cold fronts seen in Chandra images is quite remarkable. Possibly the
edges may arise from moving cold gas clouds that are the remnants of
merger activity. They may arise either from massive mergers as in
A2142, multiple collapses as suggested for RXJ1720.1+2638 (Mazzotta et
al. 2001), or gas oscillations (sloshing) in ``cooling flow'' clusters
(Markevitch et al. 2001, 2002b). For A3667, the data were of
sufficiently high quality that the parameters of the dark matter halo
associated with the observed gas cloud could be derived (Vikhlinin \&
Markevitch 2002).  Alternatively, some edges could arise from the
interaction of surviving cold, dark matter halos as they move within
the cluster potential.  High resolution, large scale structure
simulations show that dense dark matter halos, formed at very early
epochs, would not be disrupted as clusters collapse (Ghigna et
al. 1998 and Ghigna et al. 2000). While most of the dark matter halos,
having galaxy size masses, are associated with the sites of galaxy
formation, larger mass halos also may survive or may have fallen into
the cluster only recently. Hence, we might expect to find a range of
halo mass distributions moving within the cluster potential. As these
halos move, they could give rise to the multiple surface brightness
edges observed in some clusters.  Churazov et al. (2003) showed that
edges/cold fronts in cluster cores could be produced by weak shocks or
sound waves as they traverse cluster cores. Such waves can be
generated by infalling gas that provides an impetus to the core
gas. The infalling gas itself does not penetrate the core. A crude
simulation of such a scenario was used by Churazov et al. (2003) to
explain the appearance of the core of the Perseus cluster.  Still
another possible mechanism to generate edges in cluster (or group or
galaxy) atmospheres is from motions induced by buoyant bubbles of
relativistic particles produced in central AGN (Quilis et al. 2001,
Churazov et al. 2001). Buoyant bubbles can entrain, uplift and
possibly drive gas motions. Thus, a variety of physical mechanisms
have been suggested to produce the observed cold
fronts/edges. Probably, at least several of these mechanisms are
operative in different clusters and perhaps even in the same cluster
where mergers are underway and AGN in the central galaxies are
producing relativistic plasma bubbles.

\subsection{Explosive Cavities}

NGC4636 is one of the nearest and most X-ray luminous ``normal''
elliptical galaxies ($L_X \sim$2$\times 10^{41}$ ergs s$^{-1}$).  The
first X-ray imaging observations of NGC4636 from Einstein showed that,
like other luminous elliptical galaxies, NGC4636 was surrounded by an
extensive hot gas corona (Forman, Jones \& Tucker 1986).

The Chandra observation of NGC4636 shows a new phenomenon -- shocks
produced by nuclear outbursts (see Jones et al. 2002 for details of the
Chandra observation of NGC4636). The high angular
resolution
Chandra image (see Fig.~6) shows symmetric, $\sim8$
kpc long, arm-like features in the X-ray halo surrounding NGC4636.
The leading edges of these features are sharp and are accompanied by
temperature increases of $\sim30$\%, as expected from shocks
propagating in a galaxy atmosphere.

\begin{figure}
\plottwo{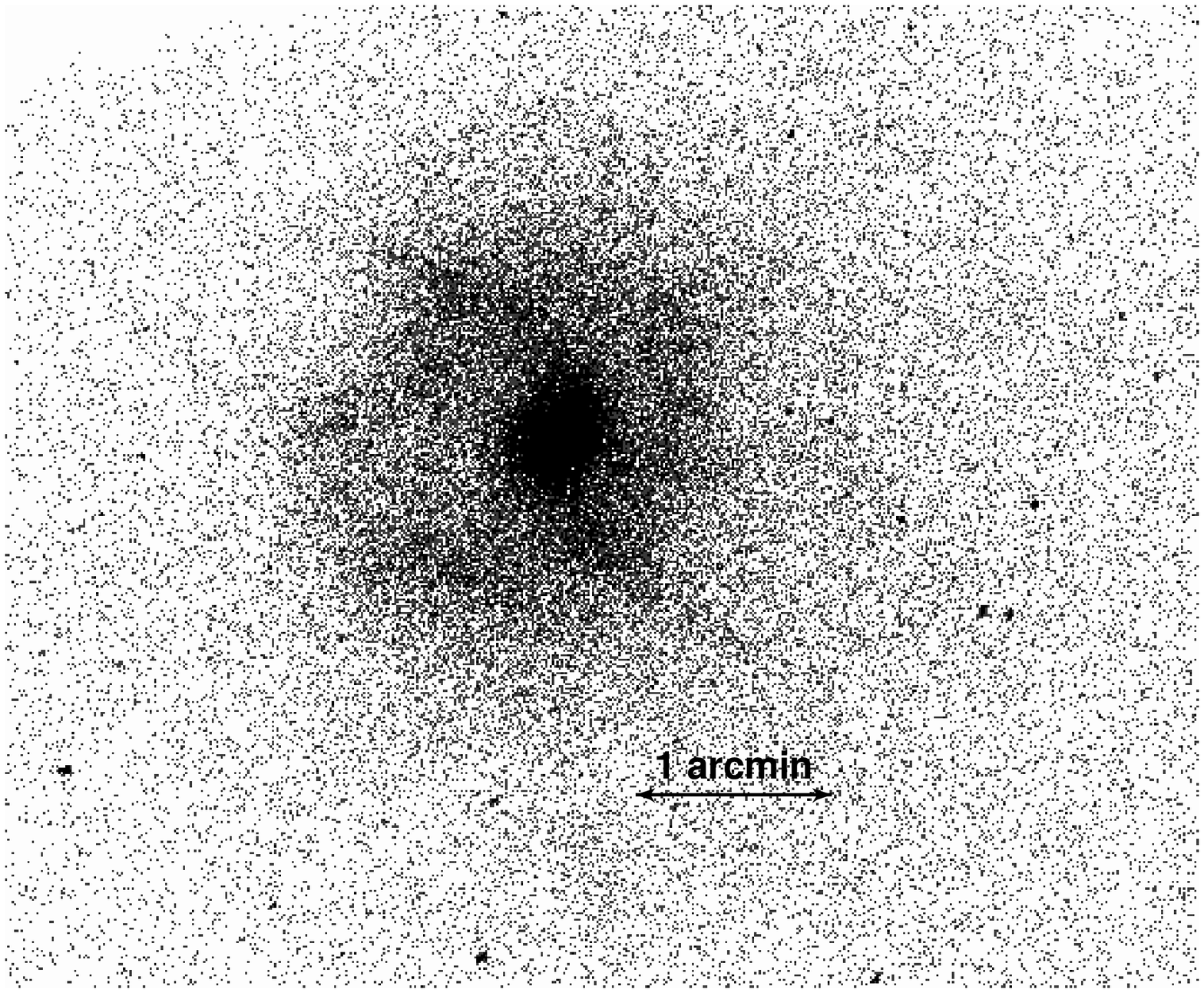}{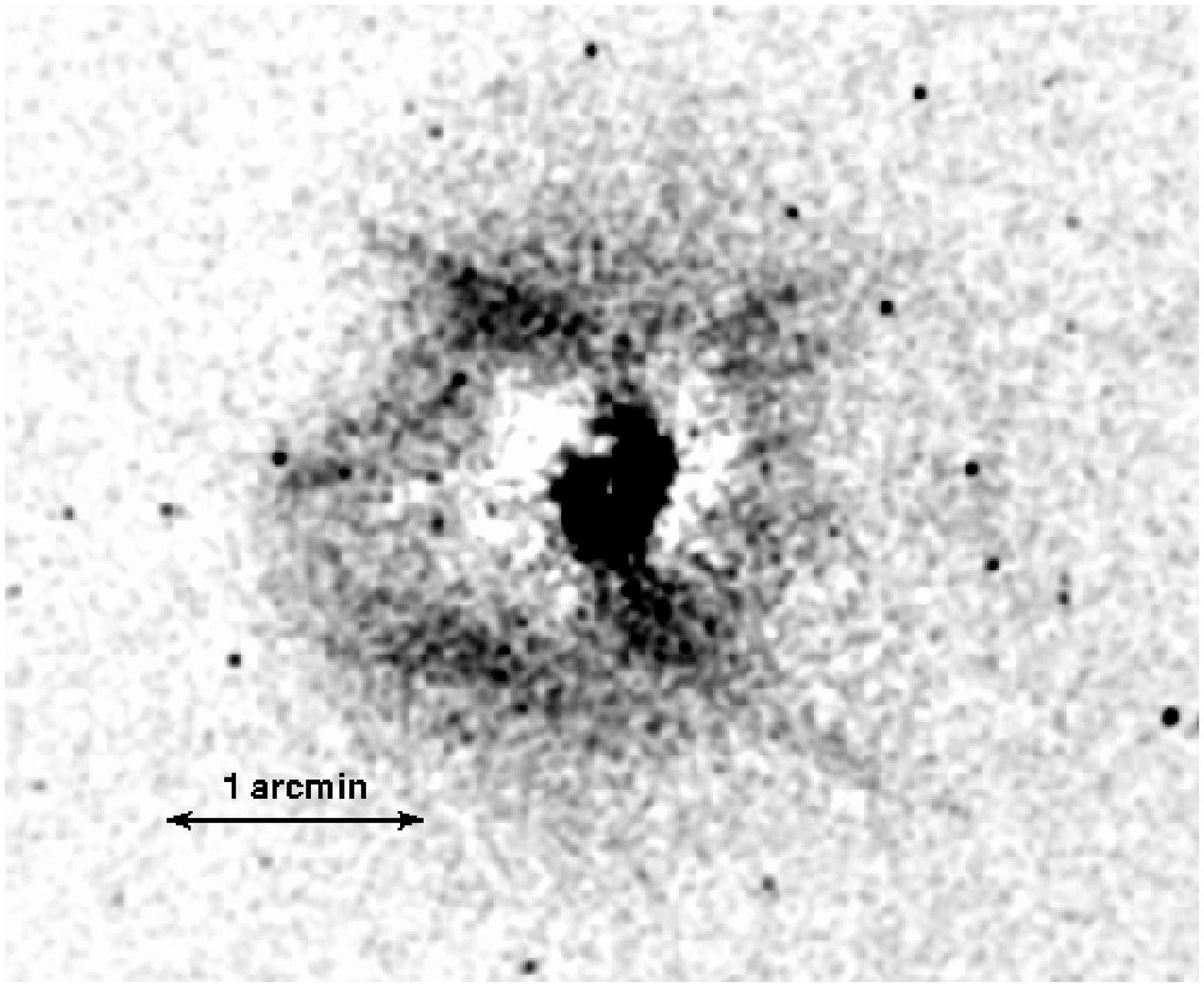}
\caption{(\textbf{a}) The 0.5-2.0 keV ACIS-S image of NGC4636. 
(\textbf{b}) The smoothed
emission after an azimuthally symmetric model describing the galaxy
corona has been subtracted.  Shocks from a recent nuclear outburst could
produce the brighter arm-like structures, while the additional
features could arise from other outbursts. }
\label{n4636_image}
\label{n4636_subrat}
\end{figure}

Although the sharpness of the edges of the NE and SW arms appears
similar to the sharp edges found along ``fronts'' in clusters (see
discussion and references above), the cluster ``fronts'' are cold,
while those in NGC4636 are hot.  Also, while the presence of sharp
fronts suggests the possibility of an ongoing merger, the east-west
symmetry of the halo structures, the similarity of this structure to
that seen around radio lobes, as well as the lack of a disturbed
morphology in the stellar core or in the stellar velocities suggest an
outburst from the nucleus as the underlying cause.  In particular, the
bright SW arm, the fainter NW arm and the bright NE arm can be
produced by the projected edges of two paraboloidal shock fronts
expanding about an east -- west axis through the nucleus.  A shock
model is also consistent with the evacuated cavities to the east and
west of the central region.

The size, symmetry, and gas density and temperature profiles of the
shocks are consistent with a nuclear outburst of energy $\sim 6 \times
10^{56}$ ergs having occurred about $\sim 3\times10^6$ years ago.  It
is tempting to suggest that these outbursts are part of a cycle in
which cooling gas fuels nuclear outbursts that periodically reheat the
cooling gas. Such outbursts if sufficiently frequent could prevent the
accumulation of significant amounts of cooled gas in the galaxy
center.

\section{Conclusions}

We have described several of the new phenomena observed with Chandra
and XMM-Newton.  Of particular note are the density ``edges'' (some of
which are shock fronts but most of which are cold fronts) and
bubbles. Some cold fronts are certainly produced through
mergers. However, others may arise from residual gas ``sloshing'' in
cluster cores (Markevitch et al. 2002). The gas motions may be
initiated by rising buoyant bubbles, weak sound waves traversing the
cluster core, or motions of dark matter halos that have survived
mergers. Buoyant bubbles of relativistic plasma can induce complex
morphologies in cluster (and group and galaxy) atmospheres as well as
moderate the effects of radiative cooling.  We have observed at least
one example of an outburst (NGC4636) with direct heating of the
surrounding gaseous atmosphere.  Clearly, cluster cores are more
complex than previous lower angular resolution observations had shown.
Studies of this ``complexity'' provide a unique opportunity to better
understand the details of cluster mergers and to investigate the
interaction of relativistic plasma produced by AGN with the gas in the
cluster cores.  

We acknowledge support from  NASA grant NAG5-9217 and NASA contract NAS8-39073.

\end{document}